\documentclass [10pt]{article}
\title{Relativistically Rotating Frames and Non-time-orthogonality}
\author{Robert D. Klauber\\1100 University Manor Dr., 38B, Fairfield, IA 52556, USA\\rklauber@netscape.net}
\date{Jan 1999}
\usepackage {graphicx}
\usepackage {longtable}
\begin{document}
\maketitle

(Published in \textit{Am. J. Phys.} \textbf{67}(2), 158-159 (Feb 99) under title "Comments 
regarding recentarticles articles on relativistically rotating frames", and 
followed therein by a response from T.A. Weber [1].)

\begin{abstract}

This paper is a brief overview of a more extensive article recently 
published in Found. Phys. Letts. [2]. Apparent disagreement with experiment 
as well as internal inconsistencies found in the traditional analysis of 
relativistically rotating frames/disks are summarized. As one example, a 
point p at 0 degrees on the circumference of a rotating disk does not, 
according to the standard theory, exist at the same moment in time as the 
same point p at 360 degrees. This and other problems with the standard 
theory are completely resolved by a novel analysis that directly addresses, 
apparently for the first time, the non-time-orthogonal nature of rotating 
frames. Though ultimately consonant with the special and general theories of 
relativity, due to non-time-orthogonality, the analysis predicts several 
peculiar (i.e., not traditionally relativistic) results. For example, the 
local circumferential speed of light is not invariant (thereby agreeing with 
the Sagnac experiment), and no Lorentz contraction exists along the disk 
rim. Other experimental results, including time dilation, mass-energy 
dependence on speed, and what has heretofore been considered a "spurious" 
signal in the most accurate Michelson-Morley experiment performed to date, 
are accurately predicted. Further, the widely accepted postulate for the 
equivalence of co-moving inertial and non-inertial rods, used liberally with 
prior rotating frame analyses, is shown to be invalid for 
non-time-orthogonal frames. This understanding of the ramifications of 
non-time-orthogonality resolves paradoxes inherent in the traditional 
theory.

\end{abstract}

\bigskip

In a recent paper on relativistically rotating disks, T.A. 
Weber\cite{Weber:1997} presents the prevailing view and appears to contend 
that one need simply apply traditional relativistic concepts directly and 
all problems and paradoxes disappear. After cordial and protracted 
communication with Professor Weber, the present writer remains convinced 
that the issue is, in fact, far from settled, and that the following 
inconsistencies remain unresolved by the standard ``solution''.

First, with regard to curvature, it is important to recognize that finite 
objects traveling geodesic paths (straight lines as seen from the lab) in 
the plane of the disk surface experience no tidal stresses, and this is true 
as seen by any observer, including those on the disk itself. Hence the disk 
surface must necessarily be Riemann flat, regardless of how one believes 
time should be defined on the disk. This is directly at odds with the 
traditional treatment.

Second, consider a continuous standard tape measure lying up against a ridge 
on the disk circumference. If we apply traditional relativity theory and 
instantaneous co-moving frames along the disk ridge, we find that the tape 
one circumference distance around the rim does not meet back up with itself 
at the same point in time. Although one may argue for local interpretation 
of standard relativity, at some point this interpretation must match up 
globally with physical reality. And a continuous tape measure which is 
temporally discontinuous can not possibly be a physical reality.

Third, in Sections V.A. and V.D. Weber reviews the traditional disk analysis 
tenet of the apparent impossibility of synchronizing a clock with itself via 
``the usual way' using light rays traveling around the disk circumference. 
But how can a coordinate system in which a clock is out of synchronization 
with itself be a reasonable representation of the real world?

In a recent article the present writer\cite{Klauber:1998} has offered a 
theoretical solution to these conundrums that agrees with all experiments. 
That paper emphasizes the following fundamental point.

Relativity theory is based on two postulates having their origin in the 
famous experiment of Michelson and Morley. These are 1) invariance of the 
speed of light, and 2) ``reference frame democracy'', i.e., all inertial 
frames are equivalent, velocity is relative. The first of these carries over 
to general relativity provided light speed measurements are made locally 
with standard rods and clocks.

The Michelson-Morley results are applicable to frames in rectilinear (not 
rotational) motion, and all of the results of relativity such as Lorentz 
contraction, time dilation, and mass-energy dependence on speed are derived 
from the two postulates based on that experiment. They are not given \textit{a priori}.

The Sagnac\cite{Post:1967} experiment, on the other hand, is a 
Michelson-Morley type experiment for rotational motion, and it showed that 
the local speed of light in a circumferential direction on rotating frames 
is not invariant\cite{Selleri:1997}. Further, it has long been known that 
all frames are not equivalent for rotational motion, as any observer can 
determine which frame is the preferred or non-rotating one (e.g., it is the 
only one without a Coriolis ``force''.)

The problem should be obvious, i.e., we can not simply assume that effects 
such as Lorentz contraction exist \textit{a priori} on the rotating disk. On the contrary, we 
have to start with new postulates based on Sagnac's results, not those of 
Michelson-Morley, and rederive relativity theory for rotating frames 
following the same steps Einstein did for rectilinear motion.

In the paper$^{{\rm 2}}$ referenced above, the writer has done just that. 
The reference frame used is the (non-Minkowskian) rotating frame itself, not 
surrogate local Minkowskian co-moving frames (which do not produce the same 
results). The analysis shows time dilation and mass-energy dependence on $v$ = 
$\omega r$, just as in standard special relativity (and therefore agreeing 
with cyclotron experiments), but no Lorentz contraction along the disk rim. 
The disk surface turns out to be Riemann flat, in agreement with tidal force 
analysis, and not curved as argued by Einstein and others. Further, a 
continuous tape measure does indeed meet back up with itself at the same 
point in time.

The lack of synchronization of a clock with itself is also resolved, since 
the underlying and tacit assumption in the ``usual way'' of synchronizing is 
Einstein's first postulate that the speed of light is invariant, i.e., the 
same in both directions around the rim. But the Sagnac experiment shows that 
this is not true, and in fact, to first order,

\begin{equation}
\label{eq1}
\vert v_{light, circumference} \vert    =   c{\rm 
} \pm   \omega r	
\end{equation}

\noindent
where the velocities in (\ref{eq1}) are \textit{physical} (not merely coordinate) values, i.e., they 
represent values which would be measured by standard physical instruments.

Further, the second relativity postulate does not apply either, as anyone 
can determine their angular velocity and their circumferential velocity 
($\omega r)$ relative to the inertial frame in which their axis of rotation 
is fixed. When light rays are used to synchronize clocks around the 
circumference by observers knowing their circumferential velocity and the 
speed of light from (\ref{eq1}) above, the synchronization turns out to be exactly 
what one finds by using light rays from a clock located at the disk center. 
Hence, a clock can be synchronized with itself using light rays traveling 
around the circumference and there is no paradox at all.

In the paper it is also shown that the ``surrogate rods postulate'' (small 
coincident inertial and non-inertial standard rods with zero relative 
velocity are equivalent), used liberally with co-moving frames in prior 
rotating disk analyses, is invalid for non-time-orthogonal frames, of which 
the rotating frame is one. In other words, Minkowski tangent frames can 
represent (curved or flat) time orthogonal frames locally, but not (curved 
or flat) non-time-orthogonal frames. This important fact appears never to 
have been realized before. As a corollary, this conclusion is true even in 
the large radius, small rotational velocity limit.

The derivation of all of these results is remarkably straightforward, 
provided one can put aside the unconscious predisposition toward a theory 
derived from different postulates than those shown by experiment to be 
applicable to rotating frames. 

With regard to the traditional argument that ``.. inertial frames used to 
interpret experiments are only approximate and invariably are part of a 
rotating system'', for every supposed rotating system we are in (e.g., earth 
around sun, sun around galactic center, etc) except one (earth surface 
around earth central axis), our frame is actually a free fall, or inertial, 
system and therefore Lorentzian. The only effective rotational velocity in 
that case is the earth surface velocity about its own (inertial) axis. 
Michelson and Gale\cite{Michelson:1925} did in fact measure the Sagnac 
effect for the earth's surface velocity in the 1920s. 

The most significant experiment, however, and the most accurate 
Michelson-Morley type test to date, is that of Brillet and 
Hall\cite{Brillet:1979}. They found a ``null'' effect at the $\Delta 
t$/$t $= 3X10$^{{\rm -} {\rm 1}{\rm 5}}$ level, ostensibly verifying standard 
relativity theory to high order. However, in order to obtain this result 
they were forced to subtract out a ``spurious'' and persistent signal of 
approximate amplitude 2X10$^{{\rm -} {\rm 1}{\rm 3}}$ at twice the rotation 
frequency of their apparatus. The theory developed by the present writer, in 
contradistinction to the standard theory, actually predicts just such an 
effect due to the earth surface velocity. For the Michelson-Morley test 
geometry this theory predicts a signal amplitude of 3.5X10$^{{\rm -} {\rm 
1}{\rm 3}}$. For the Brillet and Hall test geometry, however, the light 
paths are not restricted to two perpendicular paths, and the resultant 
$\Delta t$/$t$ effect is diluted. Brillet and Hall do not specify pertinent 
light path dimensions, but from the sketch of their apparatus, one could 
expect a reduction in signal of perhaps 30-50\% . This would result in a 
predicted amplitude range of 1.7-2.5X10$^{{\rm -} {\rm 1}{\rm 3}}$ and 
remarkably close agreement with the measured value.

With regard to electrodynamics, Ridgely\cite{Ridgely:1998} has recently 
used covariant constitutive equations in an elegant analysis to answer a 
troubling question cogently posed by Pellegrini and 
Swift\cite{Pellegrini:1995}. Ridgely derives electrodynamic results for the 
rotating frame itself, not the co-moving frame(s), and finds that those 
results match what one would find by simply applying Maxwell's equations and 
traditional special relativity to the co-moving frame(s).

The conclusion is this. Only with use of the rotating frame itself (and 
associated transformations and metric) can one obtain internally consistent 
results which agree with all experiments. However, for the purposes of time 
dilation, mass-energy, and momentum calculations (as the writer has shown in 
his paper) and Maxwell's' equations (as Ridgely has shown), one can get away 
with using traditional special relativity and local Minkowski co-moving 
frames. That is, in these cases Nature conspires to make both the rotating 
(non-traditional) and co-moving (traditional) frame solutions produce the 
same result for lab observers (i.e., mass-energy dependence on $\omega r$, 
electric polarization, etc.). When it comes to matters of time 
(synchronization, simultaneity), space (curvature), and 
Michelson-Morley/Sagnac type experiments, however, then analysis must be 
confined to the rotating frame itself, else the above-delineated 
inconsistencies and inexplicable ``spurious'' experimental signals 
inevitably arise.

Thus it appears that the rotating disk problem may have, at long last, been 
completely solved. According to Ridgely's and the present writer's analyses, 
no paradoxes remain and all theory matches up with the physical world as we 
know it.

Finally, and perhaps ironically, the writer's analysis actually turns out to 
be completely consonant with special relativity. That is, unlike other 
attempts to reconcile the Sagnac results, it leaves Lorentz covariance and 
all other traditionally relativistic effects for \textit{Minkowski} frames intact. Apparent 
differences, such as those described herein, manifest specifically for the 
non-Minkowskian rotating frame, and generally, are characteristic of 
non-time-orthogonal frames. That is, the underlying physics is the same, 
merely being seen from a different (time orthogonal vs non-time-orthogonal) 
point of view.

\end{document}